\documentclass{bioinfo2}
\copyrightyear{2016} \pubyear{2016}
\access{}
\appnotes{}
\usepackage{epstopdf}
\begin{document}
\firstpage{1}

\title[Protein-RNA interactions]{RBPBind: Quantitative prediction of Protein-RNA Interactions}
\author[Gaither \textit{et~al}.]{Jeff Gaither\,$^{\text{\sfb 1,}*}$, Yi-Hsuan Lin\,$^{\text{\sfb 2}}$ and Ralf Bundschuh\,$^{\text{\sfb 3,}*}$}
\address{$^{\text{\sf 1}}$Mathematical Biosciences Institute, The Ohio State University, Columbus, 43210, USA,\\
$^{\text{\sf 2}}$Department of Biochemistry, University of Toronto, Toronto, Ontario M5S 1A8, Canada and\\
$^{\text{\sf 3}}$Department of Physics, Department of Chemistry\&Biochemistry, Division of Hematology, Department of Internal Medicine, Center for RNA Biology, The Ohio State University, Columbus, 43210, USA.
}

\corresp{$^\ast$To whom correspondence should be addressed.}

\history{}

\editor{}

\abstract{\textbf{Summary:} We introduce RBPBind, a web-based tool for the
quantitative prediction of RNA-protein interactions. Given a user-specified RNA
and a protein selected from a set of several common RNA-binding proteins,
 RBPBind computes the binding curve and effective binding constant
 of the reaction in question. The server also computes the probability that, at a 
given concentration of protein, a protein molecule will bind to any particular nucleotide along the RNA.
The software for RBPBind is adapted from the Vienna RNA package, whose
  source code has been modified to accommodate the effects of single stranded
  RNA binding proteins.  RBPBind thus fully incorporates the effect of RNA
  secondary structure on protein-RNA interactions.\\
\textbf{Availability:} Our web server is available at \href{http://bioserv.mps.ohio-state.edu/RBPBind}{http://bioserv.mps.ohio-state.edu/RBPBind}\\
\textbf{Contact:} \href{bundschuh@mps.ohio-state.edu}{bundschuh@mps.ohio-state.edu}\\
}

\maketitle

\section{Introduction}
RNA-binding proteins (or RBPs for short) are the most important functionaries in the process of post-transcriptional regulation. RBPs and other proteins directed by them to particular RNA targets can exert control over the process of translation by recruiting ribosomes to an RNA or preventing their binding to the RNA, by conferring stability or targeting RNAs to rapid decay, and by directing the RNA to specific cellular compartments~\citep{glisovic2008rna}.

A fundamental question is whether an RBP will bind to an RNA at a given location. In addressing this, both the sequence of bases and the structure of the RNA can be important -- and recent technological advances have made it possible to address both. Most notably,  RNAcompete \citep{ray2009rapid} and RNA Bind-N-Seq \citep{lambert2014rna} use high-throughput approaches to determine the affinity of an RBP for every sequence of a fixed size -- e.g. AAAAAA to UUUUUU -- and also to establish structural preferences. 

However, \emph{in vivo} a protein interacts with an RNA molecule which might have any number of secondary structures; and structures favorable to protein-binding may not be favorable for base-pairing of the RNA. 

To address this complication, we introduce a server, RBPBind, which incorporates both quantitative sequence-dependent protein-binding affinity and the diversity of possible secondary structures which the RNA can present.  This enables us to compute the binding curve for an interaction which (as \emph{in vivo}) typically includes  many different secondary structures. We believe RBPBind is unique in its ability to compute the probability of binding for a given RBP with an arbitrary RNA on the basis of both experimentally-determined sequence preferences and secondary structure prediction.

RBPBind operates using a modified version of the Vienna RNA package \citep{lorenz2011vienna}, one of the more widely-used tools for computing RNA secondary structures. Our basic device is to computationally treat RNA-protein interactions as a phenomenon of secondary structure itself, which can be modeled with only minor changes to existing algorithms.

\begin{methods}
\section{Methods}
RBPBind incorporates protein-binding into secondary structure prediction using the paradigm of~\citep{forties2010modeling}.  Specifically, it computes the probability that an RBP will bind to an RNA through the use of \emph{partition functions}, i.e.
\begin{align}\label{maineqn}
\textbf{P}(\text{RBP binds to RNA})=\frac{\displaystyle\sum\limits_{\substack{\text{structures $U$} \\ \text{including an RBP}}}e^{-\beta \text{E}(U)}}{\displaystyle\sum\limits_{\text{all structures $S$}}e^{-\beta \text{E}(S)}}
\end{align}
where $E(U)$ is the energy of a particular protein-RNA structure, and by convention $\beta=\frac{1}{k_{B}T}$ where $k_{B}$ is Boltzmann's constant. In spite of the fact that the number of possible structures on both the top and bottom of Eq.~(\ref{maineqn}) grows exponentially with the number of bases, the Vienna RNA package can compute such sums in the absence of proteins in polynomial time, using the recursive approach set forth in~\citep{mccaskill1990equilibrium}.

To incorproate RBPs, we express our partition function (the denominator of Eq.~(\ref{maineqn})) as
\begin{align}\label{Zc}
Z(c)=Z_{0}+\sum_{i}Z_{i}\frac{c}{K_{d,i}^{(0)}}+\sum_{i<j}Z_{ij}\frac{c}{K_{d,i}^{(0)}}\frac{c}{K_{d,j}^{(0)}} + \dots
\end{align}
We assume that the protein binds to a fixed number of bases, say $m$, which then become unavailable for base-painring; this motivates our use of the quantity $Z_{i}$, which is the sum over all RNA structures in which  nucleotides $i,\ldots, i+m-1$ are unpaired. We use $Z_{0}$ to represent the partition-function sum over all structures which have \emph{no} protein attached. 

The constant $K_{d,i}$ is called the \emph{bare dissociation constant} at nucleotide $i$, and quantifies the RBP's  tendency to bind to the nucleotides $i$ through $i+m-1$ (provided these bases are unpaired).    We also note that the numerator of (\ref{maineqn}) is given by $Z(c)-Z_{0}$, allowing us to evaluate the whole expression provided we have $Z(c)$.  It was shown in~\citep{forties2010modeling} that this method of incorporating protein-binding into the partition function does not alter the $O(N^3)$ complexity of the calculation.

To arrive at $Z(c)$, however, we must obtain the bare dissociation constants $K_{d,i}^{(0)}$. Our methology in this task relies on \emph{relative} dissociation constants derived from RNAcomplete experiments. Specifically, we obtain from the RNAcompete experiments a set of measurements $K_{d,i}^{(\text{rel})}$ such that
\begin{align*}
\gamma K_{d,i}^{(\text{rel})} = K_{d,i}^{(0)}
\end{align*}
for a value $\gamma$ that is the same for every $i$. This allows us to evaluate the quantity $Z(\gamma c)$ for any $c$, using (\ref{Zc}) with $K_{d,i}^{(\text{rel})}$ in place of $K_{d,i}^{(0)}.$ To obtain the scaling factor $\gamma$,  we then solve the equation
\begin{align*}
\frac{Z(\gamma c) - Z_{0}}{Z(\gamma c)}=\frac{1}{2}
\end{align*}
for specific RNA-protein reactions whose \emph{effective dissociation constants} have been established through \textit{in vitro} experiments in the literature. Our solution $\gamma c$ then in principle equals the published effective dissociation-constants $K_{d,\text{eff}}$, and we perform a least squares procedure over several such reactions to obtain a final $\gamma$ (see Supplement for more details.)

\end{methods}

\section{Interface}
The interface of RBPBind is straightforward. The user enters an RNA sequence of up to 250 bases, and chooses one of four available proteins. (These four proteins -- HuR, RBFOX1, U2AF2 and KHDRBS3 -- were selected based on the criterion that the two required data-types, relative binding constants from RNAcompete data and independent biochemical evidence for converting these into absolute binding constants as above, were both available.)

 \begin{figure}[!tpb]
 \includegraphics[width=3.3in]{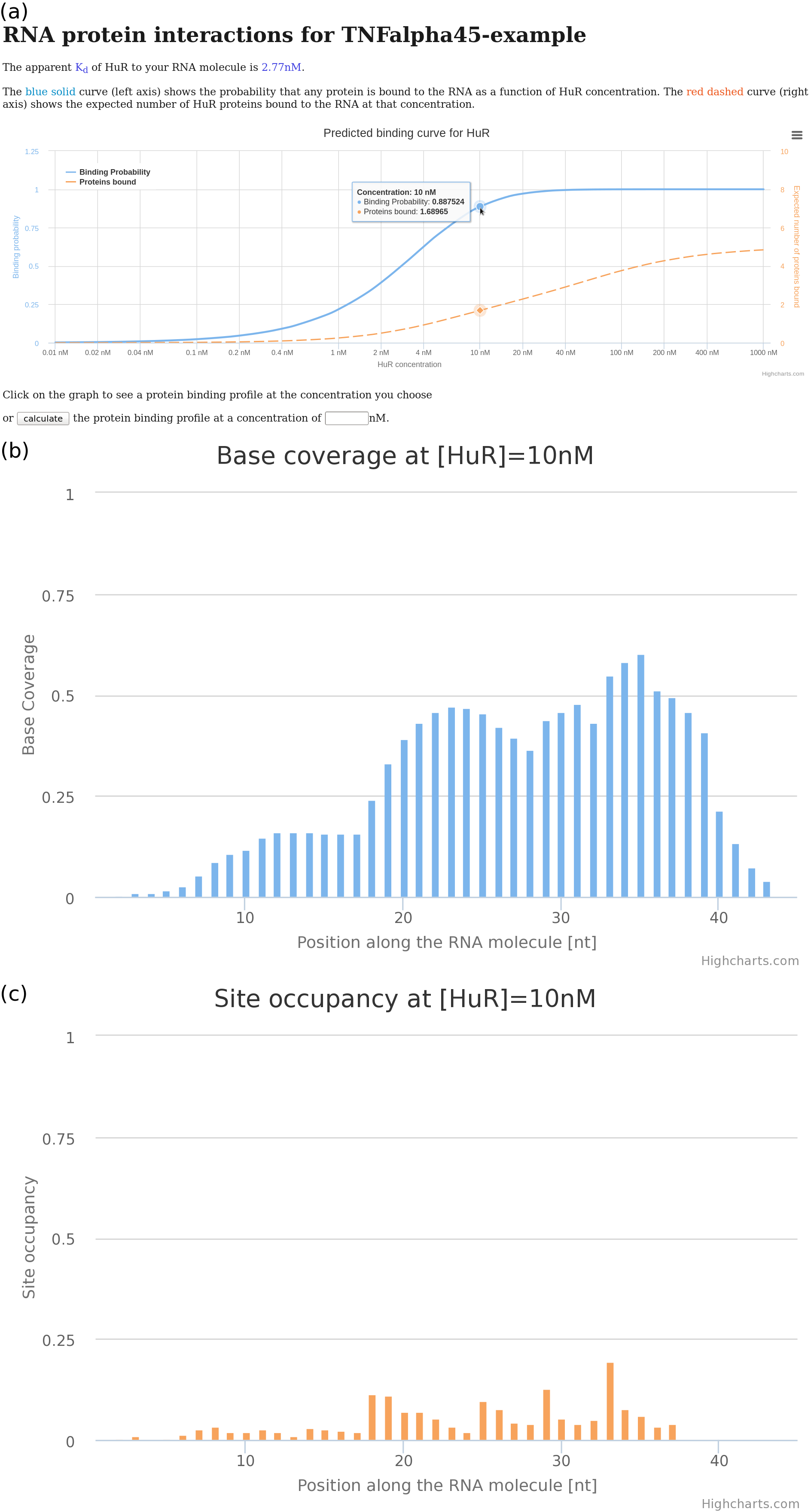}
 \caption{Output of the RBPBind web server. When provided with an RNA sequence and choice of an RNA-binding protein, the web server initially responds with the effective $K_d$ for the RNA-potein interaction, the binding curve, and the expected number of proteins bound to the RNA as a function of protein concentration (a).  After clicking on a point in the graph or manually entering a protein concentration, the sever further reports the probability that each base along the RNA is covered by a protein (b) and is the first base in the footprint of a protein (c).}\label{fig:01}
 \end{figure}

The user submits the RNA and protein, and the server opens a window with the information shown in Fig.~\ref{fig:01}(a). This output window depicts the binding curve (the probability of an RBP being bound to the RNA as a function of protein concentration), as well as expected number of proteins bound to the RNA. The effective binding constant for the protein-RNA interaction is calculated as the concentration at which half of the RNA molecules are bound by protein.

RBPBind also computes information about site-specific binding. Clicking a point on the graph, or entering a concentration in the box below it, will bring up a site-by-site profile, which plots the probability that protein will bind to  any specific base. The plot in Fig.~\ref{fig:01}(b) gives the probability that the site will be \emph{covered} by protein, i.e., that this site will lie anywhere in the footprint of a binding protein; the plot in Fig.~\ref{fig:01}(c) gives the probability that it will be \emph{occupied} by protein, i.e., that this site is the one in the protein's footprint that is closest to the 5' end of the RNA.

\section{Conclusion}

We have introduced a server which provides researchers with a predicted affinity of any RNA for a set of single-stranded RNA binding proteins taking full account of all sites on the RNA as well as the protein's competition with the RNA's secondary structure. The server provides both the strength of the interaction and the expected binding locations. Given the importance of protein-RNA interactions we expect it to be of broad use for the RNA biology community.  We anticipate adding support for additional proteins as more biochemical data becomes available.  We also plan to extend our server to allow prediction of the simultaneous interactions of multiple proteins with an RNA, as well as to cover the case of double-stranded RBPs.

\section*{Acknowledgements}
We are grateful for useful discussions with Quaid Morris from the University of Toronto and for his kindness in sharing raw RNAcompete data and analysis code with us.

\section*{Funding}

This material is based upon work supported by the National Science Foundation under Grants No. DMS-0931642 (JG) and DMR-1410172 (RB).

\bibliographystyle{natbib}

\bibliography{mybib}

\end{document}


\maketitle
\section{Obtaining relative dissociation constants}

RNAcompete protein binding preference data is reported after several
normalization and preprocessing steps described in the supplementary
information of~\citep{ray2013compendium}.  Since the final step of
calculating z-scores perturbs the interpretation of the data as
effective binding constants, the authors of~\citep{ray2013compendium}
provided us with their intermediate data. We then took our relative
dissociation constants $(K_{d,i}^{(\text{rel})})^{-1}$ as the
quantities obtained in the data-processing procedure described in the
supplementary information of~\citep{ray2013compendium} immediately
before the final column z-score calculation.  We set the affinity equal to 0.001 for all 7-mers for which it was less, since experimental error made it impossible to accurately measure affinity below a certain degree.

 To compute a weight for the 7-mers GCTCTTC and GAAGAGC, for which we have no readings since these were used functionally to synthesize the RNA pool, we use the trimmed log average over all one-base variants. For example, when considering GCTCTTC, we computed the logarithms of the affinities for ACTCTTC, CCTCTTC, $\dots$, GCTCTTT, took the average over the central 50\% and then exponentiated. 

\section{Identification of highly selective proteins}

In order to make subsequent calculations more reliable, and also because the results of RBPBind are more useful for more selective proteins, we identified the more selective of the proteins interrogated with RNAcompete. To this end, we sorted the proteins based on the proportion of 7-mers for which the affinity of the protein is $<.001$ times that of the maximal-affinity 7-mer (see Table S1.1).  Proteins for which this proportion is large (and thus the proportion of 7-mers in the range larger than 0.001 times the highest affinity sequence is small) are more selective. Ideally, we would only make use of the selective proteins; in practice, the limited amount of experimental data available to calibrate the RNAcompete measurements (see Step 3) meant that we began our search with the most selective proteins and worked our way downward. In the end our four proteins had ``trivial 7-mer'' proportions of 78\%, 58\%, 29\% and 1\%. 

\section{Obtaining effective dissociation constants}

To calibrate our relative dissociation constants $K_{d,i}^{(\text{rel})}$ into absolute dissociation constants $K_{d,i}^{(0)}$, we use measured dissociation constants for interactions of the proteins probed by RNAcompete with specific (long) RNA molecules found in the literature. Due to the uncertainty in measuring these values $K_{d,\text{eff}}$, we required a range of at least one order of magnitude among the different $K_{d,\text{eff}}$'s. Ultimately, for the protein KHDRBS3 the smallest and largest measured $K_{d,\text{eff}}$ differ by a factor of about 40 while the range is larger for the other proteins we investigated.

In locating publications containing measured affinities for protein RNA interactions, we largely rely on the Protein-RNA Interface Database (PRIDB)~\citep{lewis2011pridb}. We list the sequences of the measured RNAs and $K_{d,\text{eff}}$ values in Table S1.2. The publications from which we derive our results are: \citep{meisner2004mrna} for HuR;  \citep{auweter2006molecular} for RBFOX1; \citep{mackereth2011multi} for U2AF2; and \citep{galarneau2009star} for KHDRBS3.

\section{Determination of calibration constant}

 An effective dissociation constant $K_{d,\text{eff}}$ is the specific concentration of free protein at which exactly half the protein will be bound to RNA. It follows that, with $Z(c)$ as defined in the main paper, the quantity $K_{d,\text{eff}}$ for a particular RNA-protein reaction must satisfy
\begin{align}\label{Zeq}
\frac{Z(K_{d,\text{eff}}) - Z(0)}{Z(K_{d,\text{eff}})}=\frac{1}{2}
\end{align}
We can use this equation to solve for the needed $K_{d,i}^{(0)}$'s in the following way: For every protein, there is a fixed value $\gamma$ such that
\begin{align*}
\gamma K_{d,i}^{(\text{rel})} = K_{d,i}^{(0)}.
\end{align*}
It follows that for every $c$, we have
\begin{align*}
Z(\gamma c)=Z_{0}+\sum_{i}Z_{i}\frac{c}{K_{d,i}^{(\text{rel})}}+\sum_{i<j}Z_{ij}\frac{c}{K_{d,i}^{(\text{rel})}}\frac{c}{K_{d,j}^{(\text{rel})}} + \dots,
\end{align*}
and we can use our modified version of the Vienna RNA package~\citep{lorenz2011vienna} to resolve the above expression into an explicit number for each concentration $c$. Thus for each protein-RNA interaction, we use our modified version of the Vienna RNA package to find that value $c^{*}$  such that $\gamma c^{*}$ satisfies (\ref{Zeq}). If all the measurements and models were perfect, this would imply that $\gamma c^{*} = K_{d,\text{eff}}$. In practice, we apply this procedure over all the measured interactions available to us for a given protein, and then apply a least-squares procedure to the resulting system of equations
\begin{align*}
\log(c_{j}) + \log(\gamma) = \log(K_{d,\text{eff},j}).\;\;\;\;\;\;\;\;\; j=1\dots \# \text{ of measured interactions for current protein}.
\end{align*}
to obtain a final value for $\gamma$.

\section{Validation}

We validate our findings in two ways. First, as we do not just calibrate our RNAcompete $K_{d,i}^{\text{(rel)}}$ values for a given protein against our $K_{d,\text{eff}}$'s for that protein, but to give some perspective to those findings we attempt to calibrate \emph{all} available proteins against \emph{all} sets of effective dissociation constants. Ideally, we would find that RNAcompete data for a given protein describe the $K_{d,\text{eff}}$'s for this protein better than did the RNAcompete data for any other protein.  We summarize the results in Table S1.3. HuR has the lowest $\chi^2$ as expected; the other three proteins are 3rd, 19th and 33rd out of 202 total at explaining their own effective dissociation constants using our scheme.  We note while all rankings are in the top sixth, the ranking gets lower as the range of the experimental dissociation constants becomes smaller and the fits are overall more permissive.

Our second, and more important verification is to plot the predicted versus the actual $\log(K_{d,\text{eff}})$ for each of our four proteins. We would expect
them to cluster around the line $y=x$ if our procedure were sound, as indeed they do (see Table S1.4.)

\bibliographystyle{natbib}
\bibliography{mybib}